\documentclass[preprint,showpacs,prb]{revtex4}
\usepackage[pdftex]{graphicx}

\begin{document}

\title{Scanning gate microscopy of current-annealed single layer graphene}

\author{M. R. Connolly$^1$,K. L. Chiou$^1$,C. G. Smith$^1$,D. Anderson$^1$,G. A. C. Jones$^1$, A. Lombardo$^{2,3}$, A. Fasoli$^3$, A. C. Ferrari$^3$}
\affiliation{$^1$Cavendish Laboratory, Department of Physics, University of Cambridge, Cambridge, CB3 0HE, UK}
\affiliation{$^2$DIEET, Universita' di Palermo, Italy}
\affiliation{$^3$Department of Engineering, University of Cambridge, Cambridge, CB3 OFA, UK}%Collaboration name if desired
\date{\today}

\begin{abstract}
We have used scanning gate microscopy to explore the local conductivity of a current-annealed graphene flake. Mapping the local neutrality point (NP) after annealing at low current density reveals micron-sized inhomogeneities which reflect the temperature distribution generated by ohmic heating. Broadening of the local e-h transition is also correlated with the spatial homogeneity of the NP. Annealing at higher current density improves the NP homogeneity, but we still observe some asymmetry in the e-h conduction. We attribute this to a hole doped region close to one of the metal contacts combined with underlying striations in the NP distribution.
\end{abstract}
\maketitle

\begin{figure}[!t]
\centerline{\includegraphics[width=85mm]{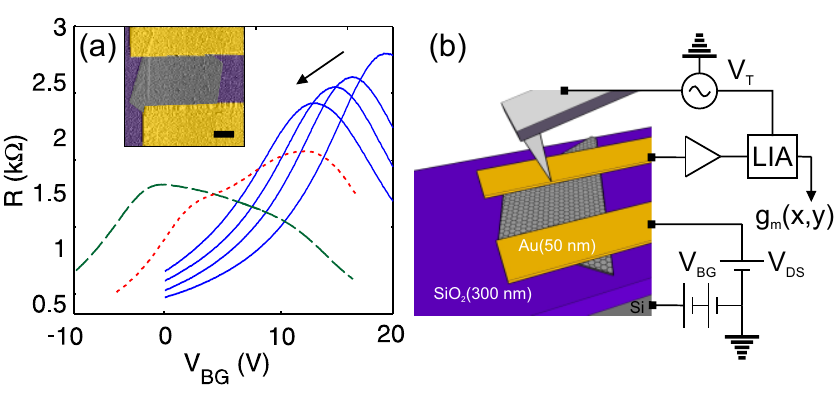}}
\caption{(a) $R(V_{BG})$ measured at 10 min intervals (blue, solid) (arrow indicates the direction of shift in peak position,) and after 9 hours (red, dotted) of current annealing at $J\approx$5$\times$10$^8$ A/cm$^2$. After 10 mins at $J\approx$1$\times$10$^9 $A/cm$^2$ (green, dashed). Inset: False-color atomic force micrograph of the sample (scale bar 2 $\mu m)$. (b) Schematic of the setup used to perform SGM.}
\label{Fig:Fig1}
\end{figure}

Graphene exhibits a wealth of properties relevant to a wide range of applications and fundamental research.\cite{Geim2007} To meet expectations and complement silicon in future nanoelectronics it is necessary to obtain precise and consistent control over its nanoscale electronic properties. Promising steps in this direction have been made using a combination of chemical functionalization and geometrical confinement.\cite{Dai2009,Han2007,AvourisPE07,KimAPL07,cervantes} Developing techniques for analyzing and controlling the effect of these processes is thus at the forefront of graphene research. Scanning probes in particular have provided valuable insights into the substrate's influence on nanometer-scale topographic and potential fluctuations.\cite{LeRoy2009,Yacoby2008,Crommie2009,Westervelt2009} At the mesoscale photocurrent\cite{Kern2008,Mueller2009} and Raman\cite{Kim2009,CasiraghiAPL,stampfer,Das2008} microscopy have also been used to characterize doping from charged surface impurities and charge transfer from metal contacts. The presence of charged impurities manifests in the electrical tranport as a shift in the charge neutrality point(NP),\cite{Novoselov2004} but the effect on the mobility and minimum conductivity at the NP is still debated.\cite{Ponomarenko2009,Fuhrer2009} Mesoscopic (micron-sized) inhomogeneities in the impurity density also contribute to the random variations observed in the transport properties of as-prepared two terminal devices.\cite{Blake2009} Annealing in an inert atmosphere and degassing reduces the NP shift and inhomogeneity,\cite{Lohmann2009,Romero2008} but unless a final cleaning step is performed \textit{in situ} the improvement is limited due to the re-adsorption of atmospheric gases and water vapor.\cite{Romero2008} Current-annealing has been used for \textit{in situ} removal of adsorbates,\cite{Moser2007,Bolotin2008} in many cases resulting in an electron-hole conduction asymmetry.\cite{Moser2007,Dai2009,andrei} Here, we use scanning gate microscopy (SGM)\cite{Crook2000} to explore the local conductivity in a current-annealed graphene monolayer and show that persistent inhomogeneities in the remanent impurity density contribute to the anomolous e-h conduction. We investigate a graphene flake ($\sim$ 8 $\times$ 8 $\mu m^{2}$) mechanically exfoliated from natural graphite onto a highly doped Si substrate capped with a 300 nm thick SiO$_{2}$ layer. Optical microscopy\cite{CasiraghiNL} and Raman spectroscopy\cite{Ferrari2006} were used to locate the flake and confirm that it is a monolayer. Two 50 nm thick Au contacts were patterned using e-beam lithography and lift-off processing [see inset of Fig. \ref{Fig:Fig1}(a).] The sample was annealed at 200$^\circ$C in N$_2$/H$_2$(5\%) to remove resist residue and mounted to the head of a scanning probe microscope evacuated to 10$^{-5}$ mbar. Further cleaning was performed \textit{in situ} by driving $J\approx$5$\times$10$^{8} $A/cm$^2$ through the device and monitoring the shift in the neutrality point voltage (V$_{NP}$).\cite{Moser2007} The two terminal resistance $R$ as a function of voltage $V_{BG}$ applied to the Si back-gate [Fig. \ref{Fig:Fig1}(a)] reveals shifts of $\approx$3 V in $V_{NP}$ every 10 minutes. We assume the initial $V_{NP}\approx$25 V is due to hole-doping adsorbates such as H$_2$O and O$_2$. Using $n$=$\alpha(V_{BG}-V_{NP})$ ($\alpha$=7.2 $\times$ 10$^{10}$ cm$^{-2}$)\cite{Novoselov2004} to estimate the carrier density, we obtain a mobility of  $\approx$8000 cm$^2$V$^{-1}$s$^{-1}$ at $n \approx$2$\times$ 10$^{11}$ cm$^{-2}$. The final cleaning current was applied for $\approx$9 hours, resulting in a shift in the bulk $V_{NP}$ from $\approx$25 V to $\approx$16 V [red curve, Fig. \ref{Fig:Fig1}(a)], and a pronounced shoulder appearing at $V_{BG}$=6 V. The latter is characteristic of flakes divided into regions with different carrier density, either by chemical doping,\cite{Farmer2009,Lohmann2009} or invasive metal contacts.\cite{Blake2009} The suppression of $R$ is also typical of flakes with a mesoscopic NP inhomogeneity.\cite{Blake2009}
Our SGM setup is shown in Fig. \ref{Fig:Fig1}(b) (see Ref.[26] for details.) To benefit from the high signal-to-noise ratio achievable using a.c. detection, we modulate the potential difference V$_T$ between the tip (Pt/Ir coated NanoWorld ARROW-NCPt) and the graphene at low frequency (typically 3 V @ 1 kHz) and detect the modulation of $I_{DS}$ ($\approx$250 $\mu$A) using a lock-in amplifier. We quantify the demodulated component ($\delta I_{DS}$$\approx$0.1 $\mu$A) by the local transconductance $ g_m =\partial I_{DS}/\partial V_T$ normalized to the bulk conductance $G(V_{BG})$.\cite{Crook2000}
\begin{figure}[!t]
\centerline{\includegraphics[width=85mm]{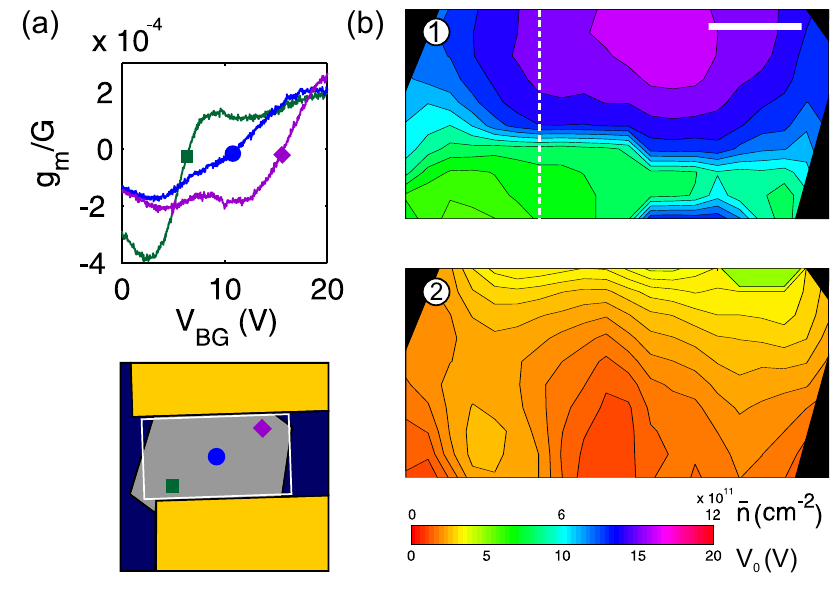}}
\caption{(a) Normalised transconductance measured with the tip at the locations marked in the inset. (b) Contour plot of the neutrality point $V_0$ of the flake in the states corresponding to the red (1) and green (2) back-gate sweeps in Fig.\ref{Fig:Fig1}(a) (Scale bar 2 $\mu$m.)}
\label{Fig:Fig2}
\end{figure}

Figure \ref{Fig:Fig2}(a) plots $g_m(V_{BG})$ measured with the tip in each of the regions indicated in the diagram of the device (inset, Fig. \ref{Fig:Fig2}.) $G$ is suppressed ($g_m<0$) at each tip position when $V_{BG}<$6 V. The opposite is true for $V_{BG}>$16 V, when $G$ is everywhere enhanced ($g_m>0$). The crossover back-gate voltage $V_0=V_{BG}$($g_m$=0) falls within the range of intermediate $V_{BG}$ and shows a sensitive dependence on tip position. To understand the basic form of the measured $g_m(V_{BG})$, we relate the conductivity $\sigma_L$ of the $\approx$(50 nm)$^2$ region perturbed by the tip\cite{Westervelt2009} to a local carrier density by $\sigma_L(x,y)=|\alpha V_{BG}+\beta V_T +\overline{n}|e\mu$, where $\beta$ is the capacitive coupling between the tip and graphene, and $\overline{n}$ is the local impurity-induced charge density.\cite{DasSarma2007, Fuhrer2009, Rossi2009} Neglecting quantum contributions to the conductivity, $g_m$ is proportional to $\partial \sigma_L(x,y)/\partial V_T$, which changes sign at the local NP when $V_{BG}=V_0\approx \overline{n}(x,y)/\alpha$ (since $\left\langle V_T\right\rangle$=0). Hence $V_0$ is a measure of $\overline{n}$, which can be empirically related to the impurity density $n_i$ via $\overline{n}\propto n_{i}^{b}$ ($b=$1.2-1.3).\cite{Fuhrer2009} We explore $\overline{n}(x,y)$ by measuring $g_{m}(V_{BG})$ with the tip positioned over a grid of points with pitch $\approx$300 nm and constructing the map of $\alpha V_0(x,y)$ shown in panel 1 of Fig. \ref{Fig:Fig2}(b). A band with $\overline{n} \approx 4.3 \times$ 10$^{11}$ cm$^{-2}$ (6 V) runs across the flake parallel to the contacts, asymetrically flanked by regions with higher density close to the contacts, where $\overline{n} \approx 11 \times$ 10$^{11}$ cm$^{-2}$ (16 V). This $\overline{n}$ profile can be explained by assuming that it reflects the temperature distribution generated by ohmic heating during current annealing. The remanent $\overline{n}$ distribution in this case suggests that energy is dissipated in the flake itself, while the contacts act as heat sinks. A similar temperature distribution was inferred from previous scanning probe\cite{Moser2007} and Raman\cite{Avouris2009} microscopy measurements, though here the asymmetry either side of the higher temperature region is more pronounced. The two values of V$_0$ also coincide with the positions of the shoulder and the maximum of the bulk $R(V_{BG})$ [red curve, Fig. \ref{Fig:Fig1}(a)], confirming that these features originate from the mesoscopic inhomogeneity.
\begin{figure}
\centerline{\includegraphics[width=90mm]{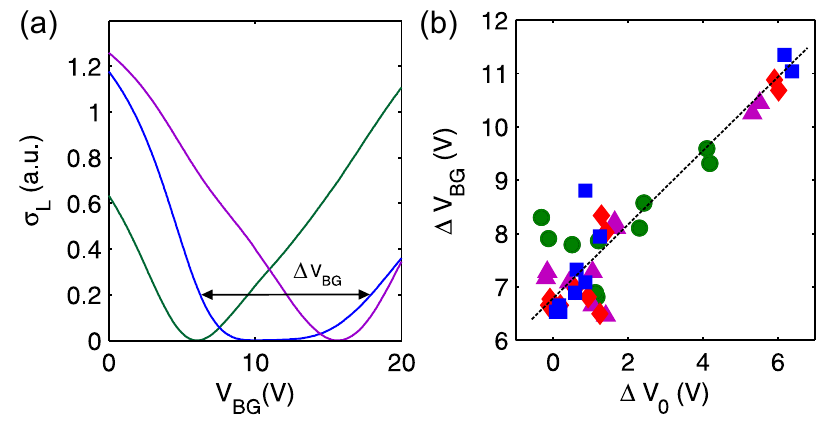}}
\caption{(a) Local conductivity at the same locations marked in Fig. \ref{Fig:Fig2}(b) (reconstructed by integrating $g_m(V_{BG})$.) Each curve has been vertically offset such that $\sigma_L(V_0)$=0. Correlation plots between $\Delta V_{BG}$ and $\Delta V_0$, the change in $V_0$ along linescans in the vicinity of the dashed line in Fig. \ref{Fig:Fig2}(b). Dashed line is drawn to indicate the linear trend.}
\label{Fig:Fig3}
\end{figure}
Figure \ref{Fig:Fig3}(a) shows plots of the local conductivity $\sigma_L(V_{BG})$ reconstructed by numerically integrating $g_m(V_{BG})$ [Fig. \ref{Fig:Fig2}(a)]. Within the framework of Ref. [30], the width of the minimum conductivity plateau $\Delta V_{BG}$ at the NP is expected to increase with $\overline{n}$.\cite{DasSarma2007,Fuhrer2009} While we did not observe this direct relationship, Fig. \ref{Fig:Fig3}(b) reveals a linear correlation between $\Delta V_{BG}$ and the spatial gradient of $V_0$ along parallel lines in the vicinity of the dashed line in Fig. \ref{Fig:Fig2}(b). (We extract $\Delta V_{BG}$ for each curve by subtracting the back-gate voltages where $\sigma_L\approx 0.2$.) This is understandable as broadening is most pronounced when the tip gates both \textit{n}- and \textit{p}-type regions, in the same way that $\Delta V_{BG}$ in homogeneous flakes is broadened by e-h puddles at low carrier density.\cite{DasSarma2007}

\begin{figure}
\centerline{\includegraphics[width=90mm]{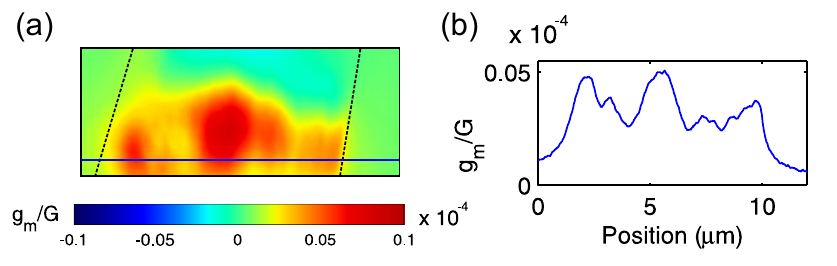}}
\caption{(a) SGM image (12 $\mu m$$\times$5 $\mu m$) mapping $g_m$ at $V_{BG}$= 3 V [c.f. green curve, Fig. \ref{Fig:Fig1}(a)]. (b) Profile of $g_m$ along the line in image (a).}
\label{Fig:Fig4}
\end{figure}

To assess the inhomogeneity remaining when $V_{NP} \approx 0$, we annealed the flake at $J \approx$1$\times$10$^9$ A/cm$^2$ for 10 minutes and repeated the measurement of $\overline{n}(x,y)$. The resulting $\overline{n}$ map is shown in panel 2 of Fig. \ref{Fig:Fig2}(b) and the corresponding $R(V_{BG})$ sweep in Fig. \ref{Fig:Fig1}(a)(green curve). In line with the previous analysis, the single peak in $R(V_{BG})$ coincides with the average value of $V_0$ ($\approx$ 2V)
($\overline{n}=1  \times$ 10$^{11}$ cm$^{-2}$), and the greater homogeneity is reflected by the absence of the shoulder. However, upon close inspection we observe a region close to the top contact with $\overline{n}=7  \times$ 10$^{11}$ cm$^{-2}$. The doping in this region, which persisted even after annealing with $J >$1$\times$10$^9$ A/cm$^2$, is responsible for the remaining asymmetry in $R(V_{BG})$, since opposing changes in $R$ occur in regions with opposite carrier type.\cite{Blake2009} Figure
\ref{Fig:Fig4}(a) shows a raw $g_m(x,y)$ ($V_{BG}$= 3 V) image of the flake after the second current anneal [c.f. green curve, Fig. \ref{Fig:Fig1}(a)]. Such images provide complementary information to the maps of $\overline{n}(x,y)$ by allowing one to resolve finer $\overline{n}$ variations directly via changes in $g_m$. 
Superimposed on the micron-sized inhomogeneity are pronounced striations in $g_m$ that span the length of the flake with lateral period of $\approx$100 nm (see linescan, Fig. \ref{Fig:Fig4}(b)). These modulations could be caused by enhanced heating from resistive hotspots in the disordered \textit{pn} junctions formed close to the metal-graphene interface,\cite{Blake2009} or electromigration of material from the contacts.\cite{Moser2007} By inspecting the amplitude of the modulation in $g_m$, we estimate that these striations reflect an impurity density fluctuation $\Delta \overline{n} \sim 7.2 \times 10^{9}$ cm$^{-2}$, which may impose an intrinsic limit to the homogeneity achievable when current annealing supported flakes.

In conclusion, SGM is a powerful method for characterizing the local conductivity of inhomogeneously doped graphene. We find that the local NP reflects the temperature distribution generated by ohmic heating and also exhibits finer linearly correlated inhomogeneities. Both types of inhomogeneity are likely to contribute to the e-h asymmetry  observed in current-annealed flakes.

This work was financially supported by the European GRAND project (ICT/FET). ACF acknowledges funding from the European Research Grant NANOPOTS and the Royal Society.

%\begin{thebibliography}{[1]}

\end{document}